\documentclass[sigconf]{acmart}
\AtBeginDocument{%
  \providecommand\BibTeX{{%
    \normalfont B\kern-0.5em{\scshape i\kern-0.25em b}\kern-0.8em\TeX}}}

\usepackage{float}
\usepackage{enumitem}
\usepackage{tcolorbox}
\usepackage{tabularx}
\usepackage{subcaption}
\usepackage{multirow}
\usepackage{colortbl}
\usepackage{array}
\usepackage[utf8]{inputenc}
\usepackage{booktabs} 
\usepackage{caption}
\usepackage{array} 
\usepackage{booktabs} 
\begin{document}







\title{I Can't Share Code, but I need Translation - An Empirical Study on Code Translation through Federated LLM}


\author{Jahnavi Kumar}
\affiliation{\textit{Research in Intelligent Software \& Human Analytics Lab}\\
Department of Computer Science and Engineering\\
Indian Institute of Technology Tirupati
\country{India}}
\email{cs22s503@iittp.ac.in}

\author{Venkata Lakshmana Sasaank Janapati}
\affiliation{\textit{Research in Intelligent Software \& Human Analytics Lab}\\
Department of Computer Science and Engineering\\
Indian Institute of Technology Tirupati
\country{India}}
\authornote{Equal Contribution}
\email{cs22b059@iittp.ac.in}

\author{Mokshith Reddy Tanguturi}
\affiliation{\textit{Research in Intelligent Software \& Human Analytics Lab}\\
Department of Computer Science and Engineering\\
Indian Institute of Technology Tirupati
\country{India}}
\authornote{Equal Contribution}
\email{cs22b046@iittp.ac.in}

\author{Sridhar Chimalakonda}
\affiliation{\textit{Research in Intelligent Software \& Human Analytics Lab}\\
Department of Computer Science and Engineering\\
Indian Institute of Technology Tirupati
\country{India}}
\email{ch@iittp.ac.in}
\renewcommand{\shortauthors}{Kumar et al.}

\begin{abstract}


Owing to the rapid evolution of technologies and project requirements, organizations need to upgrade the code base in their software projects to a new version of the programming language or even translating to an entirely new one. However, code translation is resource-intensive and requires expertise in both the source and target languages. While researchers have made progress in automating translations between legacy and modern languages, recent work has increasingly turned to pre-trained Large Language Models (LLMs) to translate efficiently.

Given the proprietary nature of code, organizations prefer fine-tuning LLMs locally rather than relying on external APIs. This is one of the first empirical studies that proposes a Federated LLM-based approach for code translation. The proposed approach enables clients to jointly train a code translator without sharing sensitive data. This study demonstrates that participants can collaboratively develop a FedLLM for efficient code translation (particularly C\# to Java and vice-versa) with superior results (more than 40\% improvement in CodeLLaMA's CodeBLEU score) compared to individual client models. Our findings indicate that FedLLM offers a collaborative approach to code translation and could serve as a promising direction for future research in this field.


\end{abstract}


\maketitle

\section{Introduction}
\label{sec: Intro}




Software migration is a critical activity adopted by industries to cater to the advances in technologies, languages, and methodologies, and to meet the demand of software evolution and maintenance \cite{fleurey2007model}. To stay relevant and efficient, companies often migrate legacy software projects to newer versions of their original programming languages or even to entirely different languages, a process known as code translation \cite{fleurey2007model, nguyen2013lexical, zhong2010mining}. Such transitions are driven by various factors: new languages may offer functionalities or optimizations that the existing language cannot provide, legacy code may be inefficient or difficult to maintain, or compatibility issues might arise with other systems that require a language update \cite{plaisted2013source}.

Translating code from one language to another is often more efficient than rewriting or re-engineering it entirely \cite{plaisted2013source}. 
However, code translation in itself is a resource-intensive and time-consuming effort, particularly for large codebases, with manual translations potentially taking years to complete \cite{oda2015learning, yang2024exploring}. For example, the Commonwealth Bank of Australia invested approximately \$750 million and dedicated five years to migrate their platform from COBOL to Java \cite{yang2024exploringunleashingpowerlarge}. These challenges have driven research efforts toward automating code translation, especially for large-scale projects \cite{tufano2019learning, lano2024using}.

Researchers have extensively leveraged Large Language Models (LLMs) for software engineering tasks, including code translation \cite{pan2024lost, yang2024exploring, hou2023large}. 
Pan et al. conducted a large-scale empirical study to investigate the capabilities of both general LLMs and code-specific LLMs for code translation across different language pairs \cite{pan2024lost}. Recently, a universal model was proposed to repair code translation errors generated by LLMs, addressing common issues such as compilation, runtime, and functional errors across multiple programming languages, and was found that the root causes of these errors are often similar, including failures to import packages, errors in loop boundaries, operator mistakes, and more \cite{yin2024rectifier}.

However, in many cases, using LLM-based translators requires companies to share their code via API with closed-source models, a concern for industries where code is a core asset and retaining privacy of code is important \cite{hou2023large}. Studies indicate that open large language models (LLMs) can achieve good performance with enhanced privacy while maintaining lower costs \cite{hanke2024open}. As a result, companies increasingly seek in-house or private LLM models trained on their proprietary data \cite{shanbhag2022exploring, kumar2024code}. In this line of work, various open-source LLMs are published that can facilitate organizations of different sizes such as LLaMA \cite{roziere2023code, touvron2023llama}, Gemma \cite{team2024gemma}, CodeT5 \cite{wang2023codet5+}, Mistral \cite{jiang2023mistral}, reducing dependence on closed-source solutions.

Open-source models benefit from the collective knowledge of multiple open-source projects. Although many companies fine-tune these models in-house, collaborative fine-tuning can be particularly advantageous for organizations within the same industry or for departments within the same organization with siloed data \cite{shanbhag2022exploring}. 
Because all the current state-of-the-art models in code translation accept code as input \cite{pan2024lost}, no existing approach provides a secure framework for collaborative training of code translators that can be trusted by proprietary entities. 

Federated Learning (FL) is one of the approaches that could be leveraged to perform code translation with two goals: (i) limit direct sharing of source code, and (ii) use data across clients while preserving privacy.  
In FL, clients or participants train a selected model on their respective datasets and then share the trained model parameters with a central server. The server aggregates these parameters to create a federated (Fed) model, which it returns to the clients \cite{bonawitz2017practical}. This federated model can either serve as the final model for the clients or act as the base for subsequent Fed rounds. In particular, FL does not require sharing private data; instead, only model parameters are exchanged, enabling a privacy-preserving collaboration mechanism \cite{bonawitz2017practical}.


In this paper, we present an empirical study that investigates a FedLLM based approach to collaboratively develop code translators using FL. The proposed approach achieves performance superior to that of individually fine-tuned client models and this highlights the benefits of collaborative training. We experimentally demonstrate this approach by focusing on the translation of code between C\# and Java using the CodeXGLUE dataset \cite{lu2021codexglue}.

Organizations often consider migrating from C\# to Java due to staffing changes or shifts in technology strategy \footnote{C\# to Java blog: \url{https://stackoverflow.com/questions/740625/migrating-a-project-from-c-sharp-to-java}}, while others migrate from Java to C\# to leverage the features of the .NET framework \footnote{Java to C\# blog: \url{https://www.eweek.com/development/migrating-from-java-to-c/}}. Our approach builds on prior research in FL for code summarization \cite{kumar2024code} and incorporates prompting techniques from recent translation studies \cite{pan2024lost}. We utilize LoRA \cite{hu2021lora} to fine-tune individual client code translation adapters, which are then aggregated using FedAvg \cite{bonawitz2017practical} to create a federated model. In contrast to the generic LLM, LLaMA2 \cite{touvron2023llama}, used in previous studies, we employ CodeLLaMA-7B \cite{roziere2023code}, a model specifically fine-tuned for code. Additionally, we investigate an alternative federated aggregation technique, FLoRA \cite{wang2024flora}, which combines clients' LoRA adapters by stacking rather than averaging them as in FedAvg.

\textit{To our knowledge, this is the first application of a Federated LLM (FedLLM) approach for code translation}, introducing a collaborative approach to privacy-preserving model training. Our findings show that the fed model offers a minimum CodeBLEU improvement of 40\% over individual client models for the chosen dataset and model.


\begin{figure}[h]
    \centering
    \includegraphics[scale=0.9]{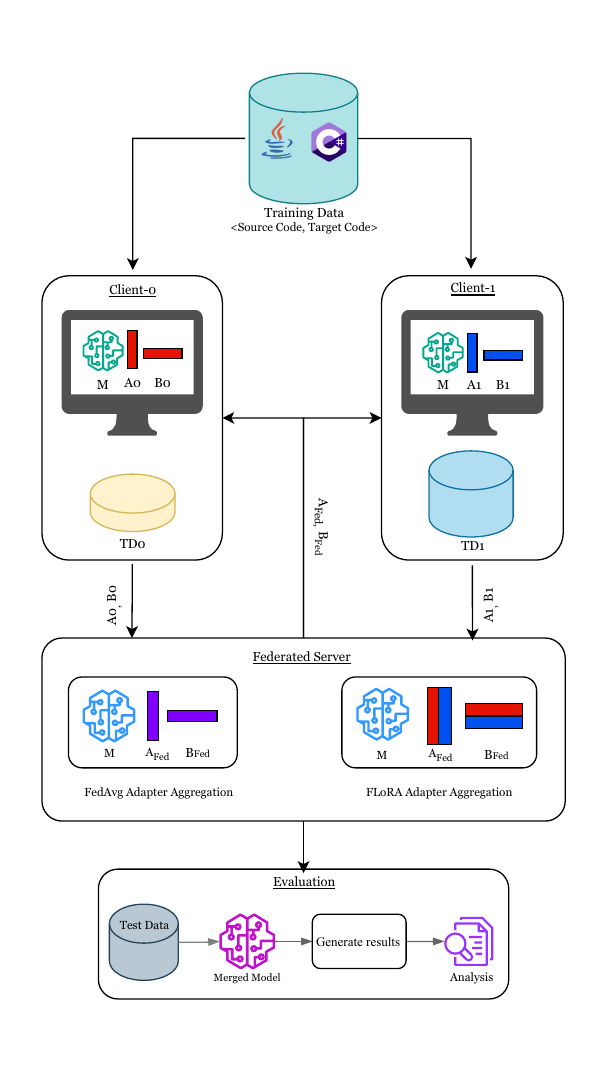}
    \caption{Overview of Code Translation Approach}
    \label{fig:MethodDiag}
\end{figure}

\section{Background}
\label{sec: bkgrnd}
Fine-tuning any pre-trained Large Language Model (LLM) for specific use cases involves significant hardware resource demands and can be time-consuming. Recent literature highlights the effectiveness of selected parameter fine-tuning, which can achieve performance comparable to that of full parameter fine-tuning \cite{h2024technical}. Among the techniques for Parameter-Efficient Fine-Tuning (PEFT), Low-Rank Adaptation (LoRA) \cite{hu2021lora} has emerged as a prominent mathematical approximation method, gaining adoptions in recent Software Engineering (SE) studies.

A recent study \cite{kumar2024code} on code summarization demonstrates that FL can effectively be applied to clients' LoRA adapters, achieving performance comparable to models trained using traditional centralized methods. This study utilizes FedAvg \cite{bonawitz2017practical}, the simplest and most widely used aggregation algorithm in FL, to aggregate the LoRA adapters from participating clients across multiple rounds. 

Equation \ref{eqn:EASE24FedAvg} illustrates the FedAvg process for client adapters \(M^a\) and \(M^b\) from clients \(a\) and \(b\), respectively. The LoRA technique decomposes the weight matrix \(M\) of size \(m \times n\) into two low-rank matrices, \(A\) and \(B\), with dimensions \(m \times r\) and \(r \times n\), where \(r\) is significantly smaller than \(n\). By applying FedAvg to the \(A\) and \(B\) matrices of all clients, we obtain server-level matrices \(A^C\) and \(B^C\), which can be matrix multiplied to reconstruct the full model weight matrix of size \(m \times n\). Therefore, the combination of LoRA and FedAvg in this software engineering study provides a foundation for employing a similar FL methodology in our research.

\begin{equation}
\label{eqn:EASE24FedAvg}
M^C = \text{Avg}(M^a, M^b) 
\end{equation}
The performance of a federated model largely depends on the aggregation technique employed for integrating updates from FL clients \cite{moshawrab2023reviewing}. Common aggregation methods include FedAvg, clipped FedAvg, homomorphic encryption, momentum aggregation, weighted aggregation, and hierarchical aggregation, among others \cite{moshawrab2023reviewing}. Specifically, in FL implementations involving Low-Rank Adaptation (LoRA), a recent study, FLoRA \cite{wang2024flora}, introduces an alternative to FedAvg. Instead of averaging, FLoRA aggregates by stacking client LoRAs.

Equation \ref{eqn:FLoRA} presents the FLoRA aggregation approach for two clients, \(a\) and \(b\). The LoRA matrices \(A^a\) and \(A^b\), each of dimension \(m \times r\) are horizontally stacked to form the server-level matrix \(A^C\) with dimensions \(m \times (r \times \#\text{clients})\). We consider the basic case where all clients agree on the same rank; however, FLoRA also supports merging clients with different rank sizes. In this example, \#clients is 2. Similarly, the matrices \(B^a\) and \(B^b\), each of dimension \(r \times n\), are vertically stacked to create \(B^C\), resulting in dimensions \((m \times \#\text{clients}) \times r\). Finally, the server-level LoRA matrices \(A^C\) and \(B^C\) are multiplied to reconstruct the full model weight matrix with dimensions \(m \times n\).
\begin{equation}
\label{eqn:FLoRA}
\begin{aligned}
A^C &= \text{horizontalStack}(A^a, A^b) \\
& \text{where } \quad A^a, A^b \in \mathbb{R}^{m \times r}, \quad A^C \in \mathbb{R}^{m \times (r \times \#\text{clients})}, \\
B^C &= \text{verticalStack}(B^a, B^b), \\
& \text{where } \quad B^a, B^b \in \mathbb{R}^{r \times n},  \quad B^C \in \mathbb{R}^{(r \times \#\text{clients}) \times n}, \\
M^C &= A^C \cdot B^C, \\
& \text{where } \quad M^C \in \mathbb{R}^{m \times n}.
\end{aligned}
\end{equation}

\section{Experimental Design}
\label{sec: exp}
This section presents the code translation dataset and standard evaluation metrics commonly used in the field. We then describe the methodology employed to train and evaluate various model types, including fed models (using both implementations discussed in Section \ref{sec: bkgrnd}), non-fed models, and individual client models on the selected dataset.
\subsection{Dataset Overview} 
The dataset used in this study is sourced from the CodeXGLUE benchmark \cite{lu2021codexglue}, a collection of datasets and tasks designed for code comprehension, generation, and translation. CodeXGLUE has been utilized in prior SE research on code translation \cite{zhang2023multilingual}, code summarization, code retrieval, and function naming \cite{ahmed2022multilingual}, among other tasks, motivating its use in this study as well. 
We utilize the \texttt{Code-to-Code Translation} dataset, which supports the task of translating code between the \texttt{Java} and \texttt{C\#} languages. This dataset comprises code snippets collected from public repositories including Lucene \footnote{Lucene: \url{https://github.com/apache/solr}}, POI \footnote{POI: \url{https://svn.apache.org/repos/asf/poi/}}, JGit \footnote{JGit: \url{ https://github.com/eclipse-jgit/jgit}}, and Antlr \footnote{Antlr: \url{ https://github.com/antlr/}}. It is divided into training, testing, and validation sets containing 10,300, 1,000, and 500 data-points respectively.

\subsection{Metrics}
This study assessed the translated code using the metrics C-BLEU \cite{papineni2002bleu}, METEOR \cite{banerjee2005meteor}, ROUGE-L \cite{lin2004rouge}, and CodeBLEU \cite{ren2020codebleu}. C-BLEU is evaluated by comparing the overlap of n-grams between the generated code and the reference code, which measures lexical similarities. METEOR, compared to the generated and reference codes, is a semantic accuracy and fluency metric that considers all parameters such as exact matches, synonyms, stemming, and word order. ROUGE-L calculates the longest common subsequence between the generated and reference codes, capturing features of structure alignment and sequence similarity. CodeBLEU is a metric that combines weighted syntactic and semantic evaluation, incorporating n-gram matching, syntax correctness, and logical consistency to assess the quality of generated code. All the metrics range from 0 to 1, and in this study, they are represented in percentages in tables. The \textit{nltk} and \textit{rouge-score} python packages are used to calculate the C-BLEU, METEOR, and ROUGE-L scores similar to the inspired study \cite{kumar2024code}. CodeBLEU was calculated using the \textit{CodeBLEU} evaluator module from the CodeXGLUE paper \cite{lu2021codexglue}.

\subsection{PEFT Modeling} 
In this research, we focus on adapting a pre-trained LLM specifically for the task of code translation. To achieve this, we employ parameter-efficient fine-tuning (PEFT) techniques, utilizing the LoRA method for our implementation. The LoRA fine-tuning targets all attention weight projections: query, value, key, and output \cite{dettmers2024qlora}. We set the LoRA module rank to 64, the highest rank the DGX system could handle without an Out-Of-Memory error. Since FLoRA creates a server with rank 128 when the client rank is 64, we chose the maximum feasible rank for stable LoRA training.

Following the fine-tuning configuration for Llama as described in recent study \cite{dettmers2024qlora}, we set a learning rate decay of ${\alpha}$ set to 16, with a maximum gradient norm of 0.3. The model operates with fp16 precision enabled. Given that the model comprises fewer than 13 billion parameters, we adjusted the LoRA dropout rate and learning rate to 0.5 and $2 \times 10^{-4}$, respectively. Throughout our experimental procedures, we utilized the Adam optimizer \cite{hu2021lora}. 

The super-vised fine-tuning and the inference prompts are adopted from a recent translation study \cite{pan2024lost} using LLMs. Following the recent FL-based code summarization methodology from \cite{kumar2024code} as the basis of our modeling approach, we construct both non-fed and fed models. To simulate federated scenarios, we model a fixed number of participants, selecting two clients for this study.

The non-fed model represents a central model where the server has full access to all clients' datasets. In this case, the LLM is trained on the merged dataset, combining data from all clients. Although central data access is often impractical in real-world scenarios due to the proprietary nature of code, we implement this non-fed model to evaluate if federated modeling can achieve comparable performance to a centrally trained model.
\begin{figure*}[h]
    \centering
    \includegraphics[scale=0.32]{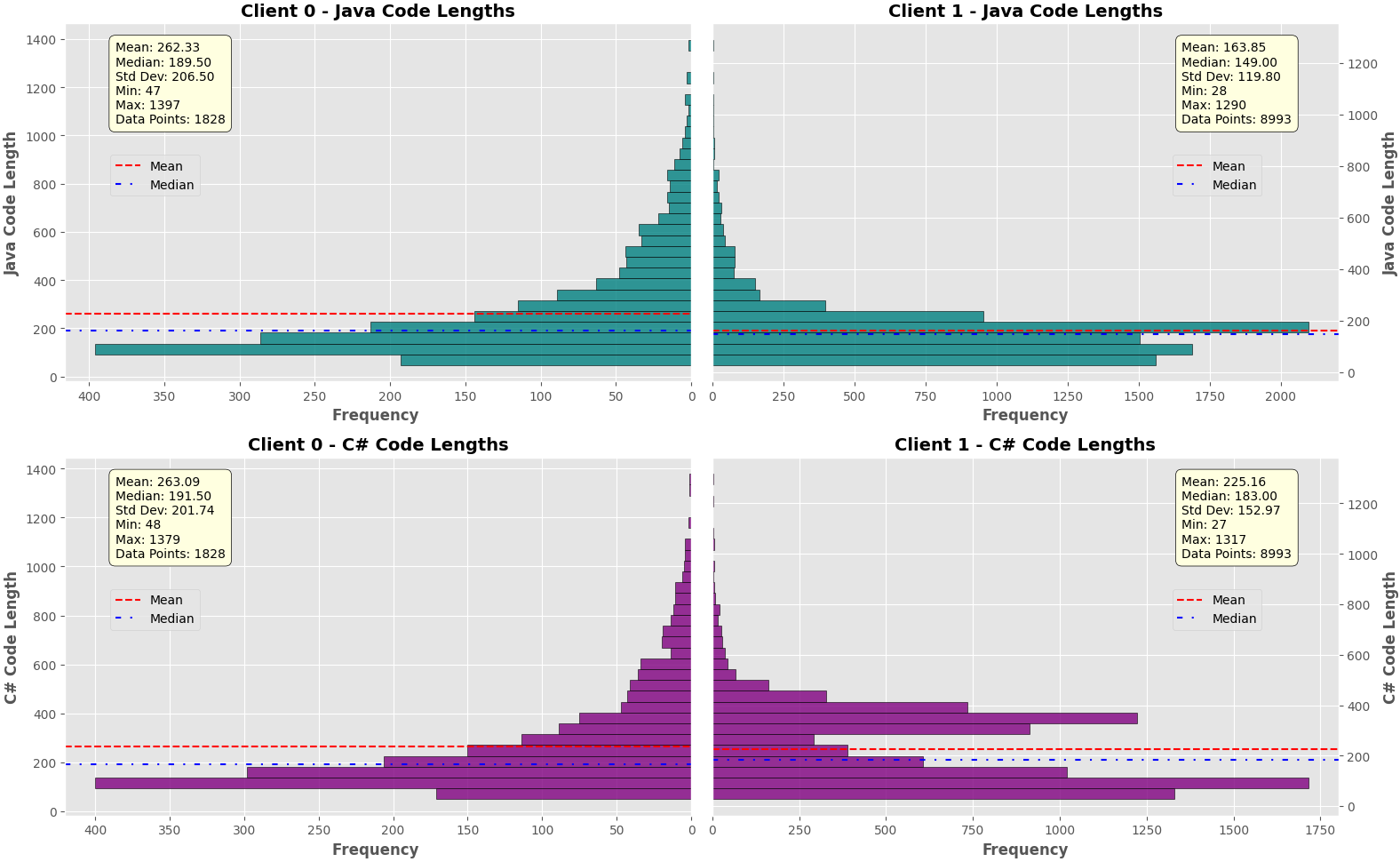}
    \caption{Data Distribution of Train set}
    \label{fig:train_DSDistri}
\end{figure*}

Federated Learning (FL), introduced in Section \ref{sec: Intro}, is applied across clients, creating a federated model in each round. In this study, we experiment with 20 FL rounds, aligned with the base study \cite{kumar2024code}, observing that model performance stabilizes within this range. However, in real-world applications, clients can adjust the number of rounds according to their performance criteria, halting training when optimal or desired performance levels are reached.

Beyond the fed and non-fed models, we also assess two additional models: the vanilla model and individual client models. The vanilla model refers to the unmodified pre-trained LLM (equivalent to the fed model at round 0), while the individual client models represent models trained independently on each client’s local dataset.

In our adaptation, we utilize 6.74B CodeLLaMA \footnote{CodeLLaMA: \url{https://huggingface.co/codellama/CodeLlama-7b-hf}}\cite{roziere2023code}, a code-focused variant of LLaMA2 \cite{touvron2023llama}, as the base LLM for all model types—vanilla, non-fed, fed, and individual-believing that a code-specialized LLM will enhance performance on code translation tasks compared to the generic LLaMA2 model \cite{touvron2023llama} used in base paper \cite{kumar2024code}.

\subsection{Dataset Preparation for Modeling}

Intrigued by the promising results from base paper \cite{kumar2024code}, a heterogeneous data distribution for clients was crafted from the CodeXGLUE dataset to enhance its suitability for federated learning simulations. Specifically, we merged the validation set with the training set to increase the training data size and then allocated this combined dataset across two clients in a 1,800 to 8,993 data-points ratio. The first client, assigned 1,800 data points, drew its samples from projects including Lucene, POI, JGit, and Antlr4, while the second client received all the data-points from the remaining projects. This allocation approach ensures a diverse and heterogeneous dataset structure suitable for FL. The test dataset consists of a separate 1,000 data points.

Figure \ref{fig:train_DSDistri} provides a detailed analysis of source code length distributions within the training dataset, segmented by client and programming language (Java and C\#). For Client0, having fewer samples, the average code length for both Java and C\# is 262 lines, with a noticeable spread indicating variability across samples. Client 1, in contrast, has a shorter average code length for Java (mean of 164) while maintaining similar distributions for C\#. These variations in code lengths across clients show us the heterogeneous nature of the dataset across clients.

\begin{figure*}[h]
    \centering
    \includegraphics[scale=0.32]{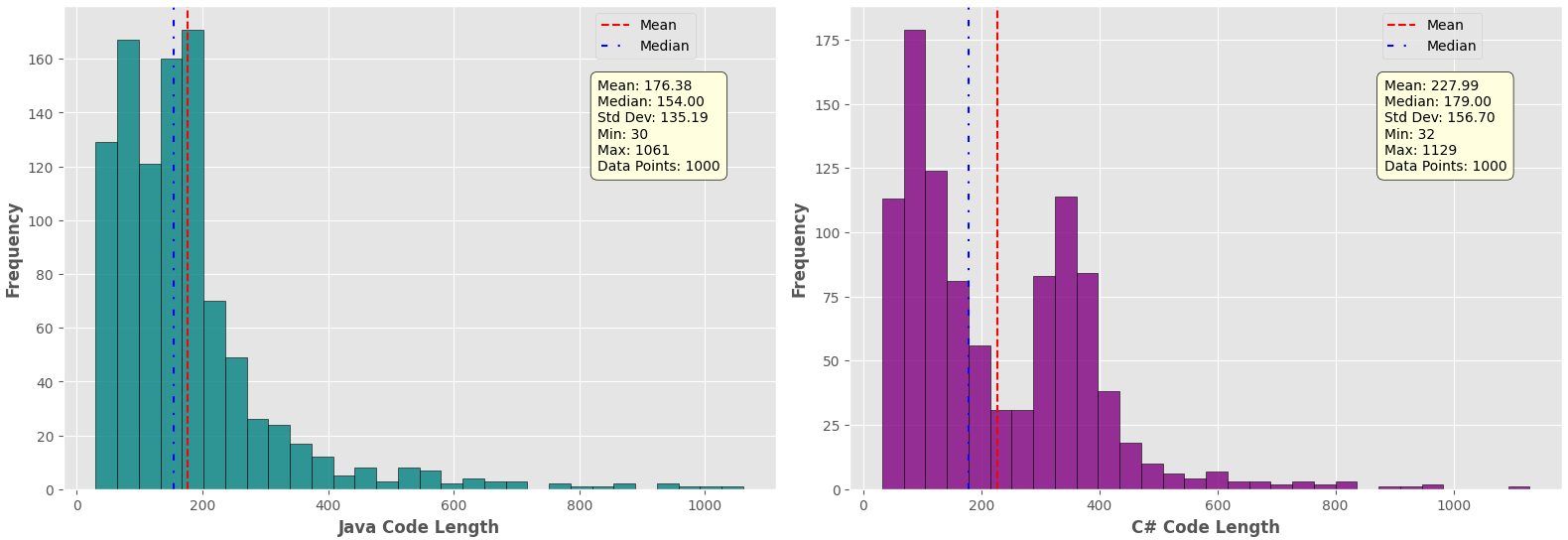}
    \caption{Data Distribution of Test set}
    \label{fig:test_DSDistri}
\end{figure*}
Figure \ref{fig:test_DSDistri} shows the code length distribution for the test dataset, presented separately for Java and C\#. The test set maintains variability, with a mean code length of 176 for Java and 228 for C\#. 

\subsection{Federated learning implementations}
We experiment with two FL + LoRA aggregation implementations introduced in Section \ref{sec: bkgrnd}: (i) Method-1 (M1): FedAvg aggregation of client LoRA adapters, a method adapted from recent code summarization research \cite{kumar2024code}, which we extend to code translation in this study; and (ii) Method-2 (M2): FLoRA \cite{wang2024flora}, a technique that stacks client LoRA adapters, which, to our knowledge, is applied here for the first time in SE research. Although LoRA rank is set to 64 for both configurations, the FLoRA server will have 128 rank because of stacking of both client's adapters.

\subsection{Overall methodology} 
This study investigates Java-to-C\# and C\#-to-Java code translation using a FedLLM framework, specifically fine-tuning CodeLLaMA with two methods: LoRA+FedAvg and FLoRA. 
An overview of these approaches is depicted in Figure \ref{fig:MethodDiag}. The full training dataset is heterogeneously divided into TD0 and TD1 for clients 0 and 1, respectively. Each client independently trains model \( M \) on their allocated dataset, generating adapters \( A \) and \( B \). The server then aggregates these adapters using FedAvg (by averaging) or FLoRA (by stacking) to create the aggregated adapters \( A_{\text{Fed}} \) and \( B_{\text{Fed}} \). These aggregated adapters are shared back to the clients for next fed round and are also subsequently merged with the model to obtain the fed model for that round, which is then evaluated for performance.

Non-federated models for both translation directions are also trained centrally, where the server has direct access to all clients' data. Additionally, the study evaluates the vanilla pre-trained LLM as well as individual models, which are fine-tuned on each client’s dataset independently. We address the following three research questions (RQs):

\noindent\textbf{RQ1: Does an individually fine-tuned model outperform the Vanilla Model?} 
This question examines whether training the pre-trained LLM on a client’s own dataset offers advantages over the unmodified vanilla model. Using a two-client scenario (see Section \ref{sec: exp}), we assess whether individual fine-tuning is beneficial for each client.
 
\noindent\textbf{RQ2: Is a collaboratively trained federated model superior to an individual model?}
This question targets the effectiveness of FL in this study: does collaborative, federated fine-tuning improve shared FL models compared to each client fine-tuning the model solely on its own data? The motivation here is to evaluate the core advantage of FL—whether collaboration through fed training benefits each client more than isolated individual training.
 
\noindent\textbf{RQ3: Can an FL-trained model (using model weight sharing) perform comparably to a centrally trained model (built with full data access)?}  
This question is sourced from the privacy-preserving nature of FL, which restricts data sharing by using model weight aggregation rather than direct data access. Although central training with full data access is likely to yield the best performance, this study explores whether fed training, while enhancing privacy, can deliver performance close to that of a centrally trained model. 

\section{Evaluation}
\label{sec: eval}
In this section, we conduct comparison and evaluation of several modeling approaches, including the vanilla model, individual client models, and federated models developed using both FedAvg and FLoRA implementations. Our evaluation employs quantitative metrics, supplemented by qualitative assessments through anecdotal comparisons of the generated results. 

\noindent\textbf{RQ1: Does an individually fine-tuned model outperform the Vanilla Model?} 

\textit{Aim}: This research question investigates whether fine-tuning a selected LLM on our dataset enhances its performance. Each client trains the pre-trained model on their respective datasets: Client 0 trains on TD0, and Client 1 trains on TD1. We evaluate performance using a common test dataset.

\textit{Expected Outcome:} We anticipate that fine-tuning the generic model on our dataset will improve its performance across all chosen metrics on the test dataset.

\textit{Actual Findings:} Despite our expectations, none of the individual models outperformed the pretrained model on the test dataset, which includes data points from both clients. This finding is significant; ideally, a fine-tuned model should generalize well to new projects (in this case, data from the other client acted as unknown projects). As shown in Table \ref{tab:RQ1_comparison}, the vanilla model consistently performed better across all metrics, indicating that the individual models lack robustness when faced with unseen data. This highlights the potential benefits of collaborative training which inspires our next RQ, where similar clients can share knowledge about code translation tasks.

\begin{table}[ht]
\centering
\sloppy
\caption{Vanilla versus Individual Model Evaluation}
\label{tab:RQ1_comparison}
\begin{tabular}{|c|l|r|r|r|r|}
\hline
\rowcolor{gray!20}
\textbf{} & \textbf{Model} & \textbf{BLEU} & \textbf{METEOR} & \textbf{ROUGE} & \textbf{CodeBLEU} \\ \hline
\multirow{3}{*}{\rotatebox{90}{Java2C\#}} & Vanilla & 5.958 & 26.278 & 28.619 & 32.601 \\ \cline{2-6}
                          & Client 0 & 0.000 & 0.300 & 0.014 & 22.889 \\ \cline{2-6}
                          & Client 1 & 0.000 & 0.108 & 0.024 & 22.058 \\ \hline
\multirow{3}{*}{\rotatebox{90}{C\#2Java}} & Vanilla & 4.402 & 29.981 & 26.791 & 42.042 \\ \cline{2-6}
                          & Client 0 & 0.000 & 0.079 & 0.011 & 22.091 \\ \cline{2-6}
                          & Client 1 & 0.000 & 0.073 & 0.020 & 21.781 \\ \hline
\end{tabular}
\end{table}

\vspace{5pt}
\noindent\colorbox{yellow!50!black}{\parbox{\dimexpr\linewidth-2\fboxsep}{\color{yellow!10!white}\textbf{RQ1 Summary}}}
\noindent\fcolorbox{yellow!50!black}{yellow!10!white}{%
    \parbox{\dimexpr\linewidth-2\fboxsep-2\fboxrule\relax}{%
        \strut Individual models did not exceed the performance of the pre-trained model on a common test dataset, indicating their lack of robustness with unseen data and suggesting potential benefits of collaborative training. \strut
    }%
}
\vspace{10pt}

\begin{figure*}[ht]
    \centering
    \begin{subfigure}{.45\linewidth}
        \centering
        \includegraphics[width=\linewidth]{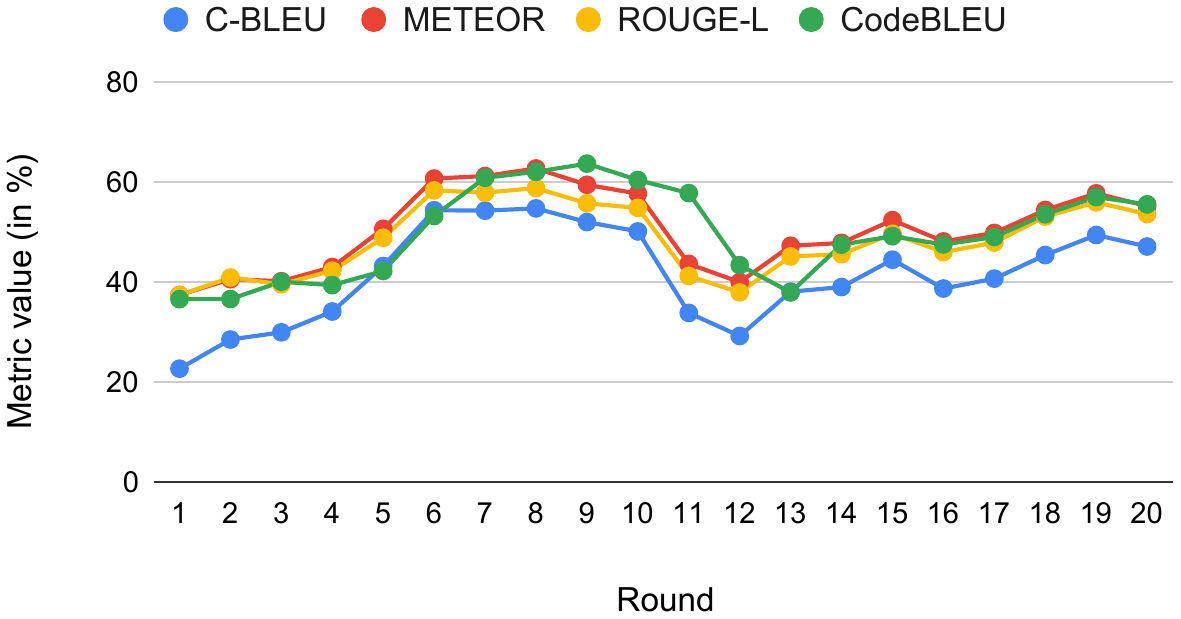}
        \caption{Java to C\# Conversion Using FedAvg}
        \label{fig:FedAvg_Java2Cs}
    \end{subfigure}
    \hfill  
    \begin{subfigure}{.45\linewidth}
        \centering
        \includegraphics[width=\linewidth]{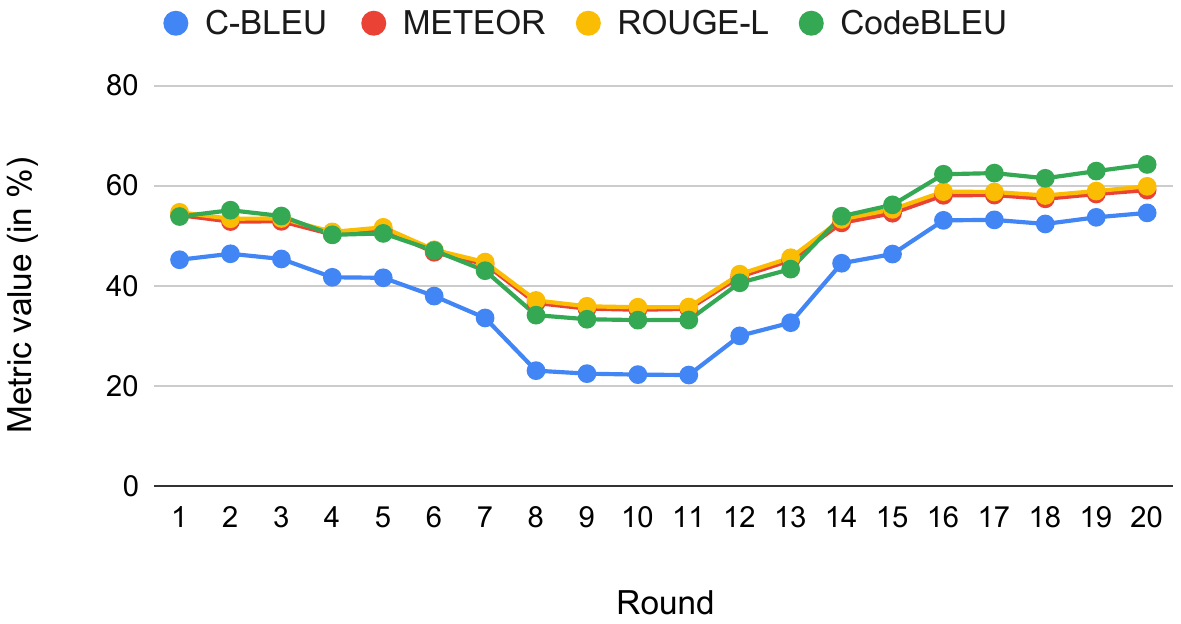}
        \caption{C\# to Java Conversion Using FedAvg}
        \label{fig:FedAvg_Cs2Java}
    \end{subfigure}
    \\[0.5cm] 
    \begin{subfigure}{.45\linewidth}
        \centering
        \includegraphics[width=\linewidth]{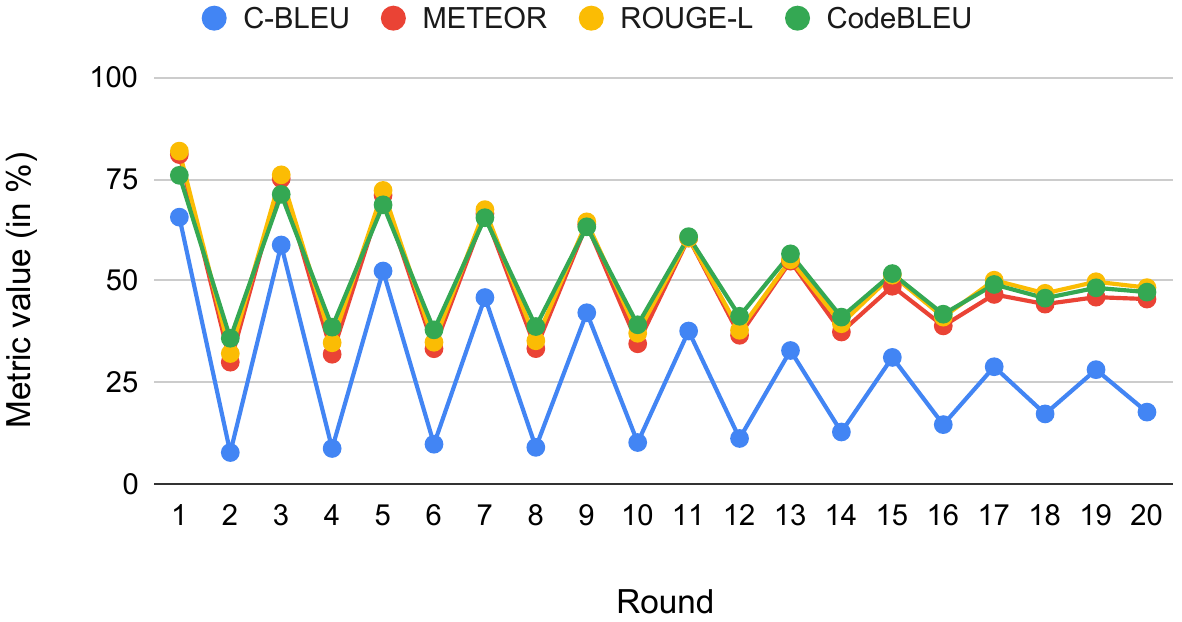}
        \caption{Java to C\# Conversion Using FLoRA}
        \label{fig:FLoRA_Java2Cs}
    \end{subfigure}
    \hfill  
    \begin{subfigure}{.45\linewidth}
        \centering
        \includegraphics[width=\linewidth]{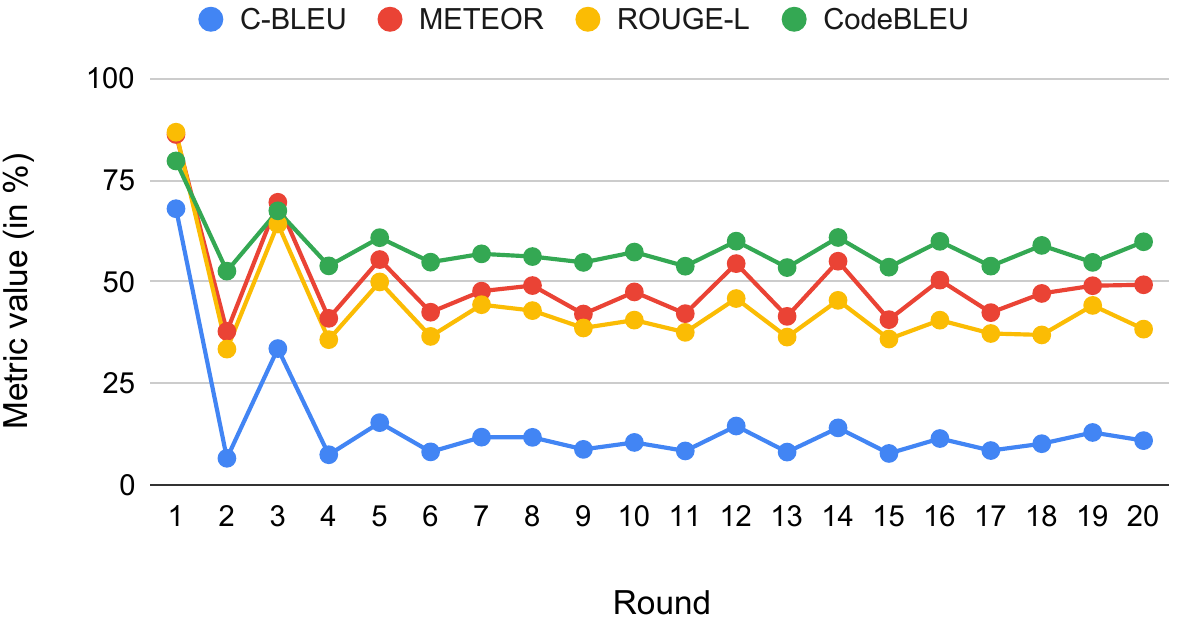}
        \caption{C\# to Java Conversion Using FLoRA}
        \label{fig:FLoRA_Cs2Java}
    \end{subfigure}
    \caption{Performance of FedLLM with FedAvg at Different Rounds}
    \label{fig:FedLLM_Rounds}
\end{figure*}
\noindent\textbf{RQ2: Is a collaboratively trained federated model superior to an individual model?}

\textit{Aim:} This RQ aims to determine whether the collaboratively trained FedLLM, developed by Client 0 and Client 1, outperforms their individual models. We investigate the benefits of collaborative training by implementing two federated learning (FL) strategies: (i) FedAvg, which averages the clients' LoRA adapters, and (ii) FLoRA, which stacks the clients' LoRA adapters. Both methods are trained over 20 rounds, with the best round selected based on the highest values of majority of performance metrics.
\begin{table}[ht]
\centering
\sloppy
\caption{Individual versus Federated Model Evaluation}
\label{tab:RQ2_comparison}
\begin{tabular}{|c|l|r|r|r|r|}
\hline
\rowcolor{gray!20}
\textbf{} & \textbf{Model} & \textbf{BLEU} & \textbf{METEOR} & \textbf{ROUGE} & \textbf{CodeBLEU} \\ \hline
\multirow{4}{*}{\rotatebox{90}{Java2C\#}} 
& Client 0 & 0.000 & 0.300 & 0.014 & 22.889 \\ \cline{2-6}
& Client 1 & 0.000 & 0.108 & 0.024 & 22.058 \\ \cline{2-6}
& \cellcolor{gray!10}FedAvg@8 & \cellcolor{gray!10}54.699 & \cellcolor{gray!10}62.675 & \cellcolor{gray!10}58.755 & \cellcolor{gray!10}62.007 \\ \cline{2-6}
& \cellcolor{gray!15}FLoRA@1 & \cellcolor{gray!15}65.594 & \cellcolor{gray!15}80.990 & \cellcolor{gray!15}81.834 & \cellcolor{gray!15}75.888 \\ \hline
\multirow{4}{*}{\rotatebox{90}{C\#2Java}} 
& Client 0 & 0.000 & 0.079 & 0.011 & 22.091 \\ \cline{2-6}
& Client 1 & 0.000 & 0.073 & 0.020 & 21.781 \\ \cline{2-6}
& \cellcolor{gray!10}FedAvg@20 & \cellcolor{gray!10}54.537 & \cellcolor{gray!10}59.093 & \cellcolor{gray!10}59.841 & \cellcolor{gray!10}64.219 \\ \cline{2-6}
& \cellcolor{gray!15}FLoRA@1 & \cellcolor{gray!15}67.945 & \cellcolor{gray!15}86.209 & \cellcolor{gray!15}86.760 & \cellcolor{gray!15}79.687 \\ \hline
\end{tabular}
\end{table}

\textit{Expected Outcome:} We anticipate that FedLLM will surpass the individual models due to: 
(i) each client training on their own dataset, enhancing performance on their specific projects; and 
(ii) the incorporation of knowledge from other participants, which should improve robustness against unseen data points.

\textit{Actual Findings:} 
For Java to C\# conversion, the optimal rounds were found to be round 8 for FedAvg and round 1 for FLoRA (see Figures \ref{fig:FedAvg_Java2Cs} and \ref{fig:FLoRA_Java2Cs}). In the C\# to Java conversion task, round 20 for FedAvg and round 1 for FLoRA yielded the best results (see Figures \ref{fig:FedAvg_Cs2Java} and \ref{fig:FLoRA_Cs2Java}). Unlike traditional FedAvg, which takes more time to converge and adapt to a heterogeneous dataset, FLoRA achieves better performance in fewer rounds by learning more effectively from heterogeneous data.

FLoRA exhibited significant fluctuations across rounds with alternating local maxima, whereas FedAvg demonstrated more stable performance with fewer maxima. Despite this variability, FLoRA achieved higher optimal performance—over 10\% better across all four metrics compared to FedAvg as observed in Table \ref{tab:RQ2_comparison}.

From Table \ref{tab:RQ2_comparison}, it is evident that FedLLM outperformed individual models by at least 50\% across BLEU, METEOR, and ROUGE metrics. For the CodeBLEU metric, FedLLM showed a minimum improvement of 40\% over individual models. Given that these metrics are derived from a common test dataset, we conclude that FedLLM is advantageous for both clients, as it facilitates learning from both their own data and peer model knowledge.

To statistically evaluate the differences in performance metrics between the FedAvg and FLoRA models at BEST round compared to individual clients, the Mann-Whitney U \cite{mann1947test} test ( with p-value 0.05) was employed. The results indicated significant differences for all comparisons, with a test statistic of 16.0 and a p-value of 0.0286 for each test. Specifically, both FedAvg and FLoRA demonstrated statistically significant improvements over Client 0 and Client 1, leading to the rejection of the null hypothesis in all cases. This suggests that both FL approaches outperform the individual client models in terms of the evaluated metrics.

\vspace{5pt}
\noindent\colorbox{yellow!50!black}{\parbox{\dimexpr\linewidth-2\fboxsep}{\color{yellow!10!white}\textbf{RQ2 Summary}}}
\noindent\fcolorbox{yellow!50!black}{yellow!10!white}{%
    \parbox{\dimexpr\linewidth-2\fboxsep-2\fboxrule\relax}{%
        \strut FedLLM exceeded the performance of individual models in code translation tasks, attaining optimal results at FedAvg round T=8 for Java to C\# and T=20 for C\# to Java, as well as T=1 for FLoRA. These results demonstrate the effectiveness of collaborative training in improving model robustness and overall performance. \strut
    }%
}
\vspace{10pt}

\noindent\textbf{RQ3: Can an FL-trained model (using model weight sharing) perform comparably to a centrally trained model (built with full data access)?} 

\textit{Aim:} This research question explores whether an FL model, which aggregates knowledge from multiple clients, can achieve performance levels comparable to a central model that has direct access to all client data. While it is understood that the central model is likely to be the most effective due to its direct data access, this question seeks to determine if the performance differences between the FL and central models are statistically significant. 

\textit{Expected Outcome:} We anticipate that the performance of the FL models FedAvg and FLoRA will not differ significantly from that of the central model. Given that the central model has access to the combined datasets TD = TD0 + TD1, it should be able to learn from diverse project data more effectively than the federated approach, which relies on a split-train-merge mechanism. 

\textit{Actual Findings:}
In both translation tasks, the central model outperformed both FedAvg and FLoRA models as shown in Table \ref{tab:RQ3_comparison}. The performance metrics exhibited similar trends across both tasks. 
\begin{table}[ht]
\centering
\sloppy
\caption{Federated versus Central Model Evaluation}
\label{tab:RQ3_comparison}
\begin{tabular}{|c|l|r|r|r|r|}
\hline
\rowcolor{gray!20}
\textbf{} & \textbf{Model} & \textbf{BLEU} & \textbf{METEOR} & \textbf{ROUGE} & \textbf{CodeBLEU} \\ \hline
\multirow{3}{*}{\rotatebox{90}{Java2C\#}} & FedAvg@8 & 54.70 & 62.68 & 58.77 & 62.01 \\ \cline{2-6}
                          & FLoRA@1 & 65.59 & 80.99 & 81.86 & 75.89 \\ \cline{2-6}
                          & Central & 75.59 & 86.94 & 87.17 & 83.18 \\ \hline
\multirow{3}{*}{\rotatebox{90}{C\#2Java}} & FedAvg@20 & 54.54 & 59.09 & 59.84 & 64.21 \\ \cline{2-6}
                          & FLoRA@1 & 67.95 & 86.21 & 86.76 & 79.69 \\ \cline{2-6}
                          & Central & 74.38 & 88.58 & 89.35 & 83.26 \\ \hline
\end{tabular}
\end{table} 

In the Java to C\# translation task, the central model demonstrated superior performance over the FedLLM models, with improvements in BLEU scores ranging from 10\% to 20\%, METEOR scores from 6\% to 24\%, ROUGE scores from 6\% to 28\%, and CodeBLEU scores from 7\% to 20\%. 

Similarly, in the C\# to Java translation task, the central model outperformed the FedLLM models, achieving increases in BLEU scores by 6\% to 20\%, METEOR scores by 2\% to 28\%, ROUGE scores by 2\% to 30\%, and CodeBLEU scores by 3\% to 18\%.
 
In summary, while FLoRA performed comparably to the central model, FedAvg exhibited greater difference in performance. To further investigate these observations, the Wilcoxon signed-rank non-parametric test \cite{woolson2007wilcoxon} was employed to statistically evaluate the performance of the FL-trained models FedAvg and FLoRA against the centrally trained model. The results indicated no significant differences across all comparisons, with a test statistic of 0.0 and a p-value of 0.125 for each test. This suggests that both the FedAvg and FLoRA models perform similarly to the central model across both translation tasks: Java to C\# and C\# to Java.

\vspace{5pt}
\noindent\colorbox{yellow!50!black}{\parbox{\dimexpr\linewidth-2\fboxsep}{\color{yellow!10!white}\textbf{RQ3 Summary}}}
\noindent\fcolorbox{yellow!50!black}{yellow!10!white}{%
    \parbox{\dimexpr\linewidth-2\fboxsep-2\fboxrule\relax}{%
        \strut The findings indicate that while the central model consistently outperforms both FL models FedAvg and FLoRA across translation tasks, FLoRA performs comparably to the central model while maintaining data privacy. \strut
    }%
}
\vspace{10pt}

\noindent{\LARGE\textbf{Anecdotal examples}}

\noindent In our comparative analysis of model outputs for migration task, we evaluated the performance of different models based on two key factors: the correctness of the translation from source to target and, in cases where translations were correct, their closeness to the provided ground truth. 

\begin{table*}[t]
\centering
\caption{Example Model Outputs for Java to C\# Migration}
\label{tab:Java2Cs_eg}
\begin{tabular}{|l|p{14cm}|}
\hline
\textbf{Java code} & 
\begin{verbatim}
public void print(String str) {
    write(str != null ? str : String.valueOf((Object) null));}
\end{verbatim}
\\
\hline 
\rowcolor{gray!20}
\textbf{Model} & \textbf{Translated C\# code} \\
\hline
\textbf{Ground Truth} & 
\begin{verbatim}
public virtual void print(string str){
    write(str != null ? str : Sharpen.StringHelper.GetValueOf((object)null));}
\end{verbatim}
\\
\hline
\textbf{Central} & 
\begin{verbatim}
public virtual void print(string str){
    write(str != null ? str : System.Convert.ToString((object)null));}
\end{verbatim}
 \\
\hline
\textbf{FLoRA} & 
\begin{verbatim}
public void Print(string str){
    Write(str != null ? str : string.Format("{0}", (object)null));}
\end{verbatim}
\\
\hline
\textbf{FedAvg} & 
\begin{verbatim}
public void print(string str){
    write(str != null ? str : Sharpen.StringHelper.GetValue((object)null));}
\end{verbatim}
 \\
\hline
\textbf{Pre-trained} & 
\begin{verbatim}
public void print(String str) {
    write(str != null ? str : String.valueOf((Object) null));}
\end{verbatim}
\\
\hline
\end{tabular}
\end{table*}
Table \ref{tab:Java2Cs_eg} shows an anecdotal example Java function and its translated C\# code generated by models. The Java function \texttt{print(String str)} is designed to output a given string to a specified destination, and to avoid null reference errors, if the input string is null, it substitutes to the string value \texttt{"null"}. The results indicate that the Central model outperforms the others, accurately translating the function while maintaining structural integrity and logical consistency with the ground truth. Following Central, the FedAvg model also delivers a correct translation but lacks the \texttt{virtual} keyword. The FLoRA model, while logically sound, utilizes \texttt{string.Format} for null handling, which deviates from the ground truth's approach. Finally, the output of the pre-trained model is identical to the Java code and does not actually translate into C\# code. 

\begin{table*}[t]
\centering
\caption{Example Model Outputs for C\# to Java Migration. Unrelated long texts generated suffixing to the actual output is shown as <unrelated\_examples\_continued>.}
\label{tab:Cs2Java_eg}
\begin{tabular}{|l|p{14cm}|}
\hline
\textbf{C\# code} & 
\begin{verbatim}
public int CompareTo(SearcherTracker other){
    return other.RecordTimeSec.CompareTo(RecordTimeSec);}
\end{verbatim}
\\
\hline 
\rowcolor{gray!20}
\textbf{Model} & \textbf{Translated Java code} \\
\hline
\textbf{Ground Truth} & 
\begin{verbatim}
public int compareTo(SearcherTracker other) {
    return Double.compare(other.recordTimeSec, recordTimeSec);}
\end{verbatim}
\\
\hline
\textbf{Central} & 
\begin{verbatim}
public int compareTo(SearcherTracker other) {
    return other.recordTimeSec.compareTo(recordTimeSec);}
\end{verbatim}
 \\
\hline
\textbf{FLoRA} & 
\begin{verbatim}
public int compareTo(SearcherTracker other) {
    return other.getRecordTimeSec().compareTo(getRecordTimeSec());}
\end{verbatim}
\\
\hline
\textbf{FedAvg} & 
\begin{verbatim}
public int CompareTo(SearcherTracker other) {
    return other.recordTimeSec.CompareTo(recordTimeSec);}
\end{verbatim}
 \\
\hline
\textbf{Pre-trained} & 
\begin{verbatim}
public int compareTo(SearcherTracker other){
    return other.recordTimeSec.compareTo(recordTimeSec);}
<unrelated_examples_continued>
\end{verbatim}
\\
\hline
\end{tabular}
\end{table*}
Table \ref{tab:Cs2Java_eg} presents an example of a C\# function 
and its Java translations generated by different models. The function is designed to compare the \texttt{RecordTimeSec} of two \texttt{SearcherTracker} instances, and return a comparison result. The Central model translates the function accurately while preserving both structural integrity and logical consistency with the ground truth. Following Central, the FedAvg model also provides a correct translation but uses inconsistent casing in the method name \texttt{CompareTo} vs. \texttt{compareTo}. The FLoRA model is logically valid, however employs getter methods \texttt{getRecordTimeSec()} for accessing \texttt{recordTimeSec}, which differs from the direct value access used in the ground truth. Lastly, the Pre-trained output generates similar to the FedAvg model's structure but continues to generate random unrelated examples. 

To summarize, we find that the translations produced by the models can be ranked in the following order: Central > FedLLM (FedAvg, FLoRA) > Pre-Trained. FLoRA model generates logically correct outputs, however they deviate from the ground truth, whereas the outputs of the FedAvg model are found to be closer to the ground truth. The outputs from the pre-trained models tend to generate generic and excessively repetitive text, often incorporating translations in other languages as well. This indicates pre-trained model struggles to accurately comprehend the specific task at hand.

\section{Discussion}
\label{sec: Discuss}
This study is a first step towards the goal of supporting privacy-preserving collaborations between industry and academia in leveraging a shared LLM fine-tuned for specific SE tasks, and in particular for the task of code translation.
For the LoRA fine-tuning, the study employs FedAvg \cite{mcmahan2017communication}, a widely used and straightforward aggregation algorithm in FL. We selected FedAvg with the rationale that if FedLLM achieves comparable results to centralized training using this basic approach, then leveraging more advanced aggregation algorithms should further improve FedLLM performance, potentially reducing the number of training rounds required. Additionally, this study also experiments with FLoRA technique \cite{wang2024flora} introduced in recent research paper, which offers a stacked aggregation of client LoRA adapters. 

LLMs are vulnerable to risks involving the unintentional exposure of sensitive information \cite{zou2023universal}. In FedLLM training, each client gains access to knowledge learned from other clients' data, which can place sensitive data, the model, and client security at risk. Consequently, FedLLM should be developed collaboratively, engaging only trusted participants \cite{arisdakessian2023coalitional}. Additionally, in real-world applications, a client might submit corrupted (erroneous) or tampered (maliciously manipulated) weights \cite{lyu2022privacy}. Hence, it is essential for the server to implement a validation mechanism to assess the trustworthiness of received weights before incorporating them into the global aggregation.

\section{Threats to Validity}
\label{sec: Threats}
\textbf{External validity: }
This study focused on training an LLM for code translation using a specific GitHub dataset, whose unique characteristics—including the languages and projects selected—may limit the generalizability of our findings. Datasets that include proprietary projects or data-points from other platforms might yield different results due to their distinct attributes. However, we have divided the data heterogeneously among clients to simulate a federated scenario that more closely resembles closed-source conditions, thus we believe that our results could be cautiously applied to closed-source training scenarios. Further research across varied datasets, models, and client configurations would contribute to strengthening the generalizability of these findings.

\noindent\textbf{Internal validity: } 
This study adapts an open-source dataset for simulated federated experiments, which may differ from real-life federated scenarios. However, to keep it closer to the real-world conditions, we mined project information from the latest stable version of each selected project and divided data points heterogeneously among clients based on this information.

The training process in this study is subject to limitations, including the nature of the training dataset (which offers a single ground truth for reference, even though code can be written in multiple valid ways in any language), as well as the chosen metrics, number of epochs, LoRA hyper-parameters, model architecture, fine-tuning methods, and federated aggregation techniques. These factors could influence the optimal number of federated rounds and the findings \cite{jongeling2015choosing}. Further empirical research is necessary to develop more effective code translation models.

\noindent\textbf{Reliability: } 
Reliability assesses the consistency of results when experiments are repeated with same data. Since no prior studies exist on FedLLM training for code translation, we are unable to validate our findings. Our results are based on our implementation of FedLLM with LoRA, and future research efforts could consider replicating this work across diverse federated architectures, as differences in design could influence both the efficiency and outcomes of the experiments.

\noindent\textbf{Construct validity: }
The effectiveness of a fine-tuned model for code translation depends on the optimal hyper-parameter configuration. To maximize the performance of the selected LLM in this task, we choose the highest rank and target modules  from recent study \cite{kumar2024code} based on our hardware's configuration, and set the other hyper-parameters such as learning rate as per the literature \cite{dettmers2024qlora}.

\section{Related Work}
\label{sec: Related}
\vspace{5pt}
\noindent\textbf{Code Translation:}

\noindent As software projects evolve, they frequently undergo various code-to-code transformations, such as software restructuring, forward engineering, platform migration, code reuse, and language migration \cite{cordy2002source}. The language migration are often driven by the enhanced functionalities, packages, or compatibility that a different programming language may offer, aligning better with evolving project requirements and specific design decisions \cite{plaisted2013source, fleurey2007model, nguyen2013lexical, zhong2010mining}. Migrating a project to a new language, when needed, is typically achieved more reliably and cost-effectively through code translation rather than a complete rewrite of the codebase \cite{plaisted2013source}. However, manual source-to-source translation can require years of effort for large projects, making it a resource-intensive task \cite{oda2015learning, yang2024exploring}.

In response, recent years have seen considerable research interest in developing automated code translators \cite{tufano2019learning, lano2024using}. These efforts have explored diverse approaches, including the use of abstract intermediate languages or library-like constructs to facilitate translation into a target language \cite{plaisted2013source}. Additionally, preserving the original source file structure during translation has been shown to aid in maintaining the translated codebase independently of the original source \cite{andrews1996macro}. 

Early work on automated translation primarily relied on rule-based systems, employing techniques like tree transformation languages \cite{cordy2002source} and tools such as the Gentle compiler construction system \cite{schroer1997gentle}. Over time, however, these approaches have shifted toward model-based methods, which leverage machine learning to capture complex linguistic patterns and structures within the source code and generate high-quality target code.  

\vspace{5pt}
\noindent\textbf{Models for Code Translation:} 

\noindent The evolution of natural language processing techniques in code translation has progressed significantly, moving from statistical machine translation \cite{oda2015learning} and neural networks \cite{wang2019domain} to large pre-trained models \cite{pan2024lost}. LLMs, which are pre-trained on vast datasets, have demonstrated strong performance on general tasks and can be further fine-tuned to excel in specific domains, including SE applications such as code translation \cite{pan2023stelocoder, pan2024lost}. 

In recent studies, LLMs have been assessed for their effectiveness and robustness in code migration tasks. For instance, a study \cite{yang2023assessing} focus on adversarial robustness in Python-to-Java and Java-to-Python migration, using models like CodeT5 \cite{wang2021codet5}, CodeBERT \cite{feng2020codebert}, CodeGPT \cite{lu2021codexglue}, and CodeGen \cite{nijkamp2022codegen}. Another study \cite{pan2024lost} analyze translation errors introduced by several LLMs, including CodeGen, StarCoder \cite{li2023starcoder}, GPT-4 \cite{achiam2023gpt}, and LLaMA-2. Translation frameworks like UniTrans \cite{yang2024exploring} augment these efforts by proposing test case augmentation for improving translation, utilizing LLMs such as GPT-3.5, LLaMA, CodeGen, and TransCoder \cite{lachaux2020unsupervised}. Further, SteloCoder \cite{pan2023stelocoder} explores Mixture-of-Experts architectures for multi-language translation into Python, offering specialization techniques for code migration.

Building on these recent advancements \cite{dhruv2024leveraging, di2024codefuse}, our study adapts CodeLLaMA \cite{roziere2023code}, an LLM pre-trained on multiple programming languages, as the foundational model. This adaptation is inspired by recent empirical evaluations of LLMs for code translation \cite{pan2024lost}, which provide insights into model fine-tuning and inference strategies. Unlike these traditional centralized approaches, we further enhance this methodology into another dimension by training the model using FL to maintain data privacy, following the approach outlined in the code summarization work \cite{kumar2024code}.

\vspace{5pt}
\noindent\textbf{Federated Learning for Software Engineering:}

\noindent Federated Learning decentralizes model training by sharing only locally trained model weights with a central server, preserving data privacy by avoiding direct data sharing \cite{bonawitz2017practical}. Several FL frameworks, including Tensorflow Federated, PySyft, FATE, Flower, PaddleFL, and Fedlearner, are now emerging to support FL implementation across various safety-critical applications \cite{du2023understanding}. While studies reveal that FL’s common challenges, such as client-server communication issues, can impact performance, we employ a simulated FL environment in this study. This approach mitigates communication, latency, and system limitations, focusing instead on determining FL's effectiveness in code translation. Practical, real-world FL implementations may follow these simulated studies to validate effectiveness under real client conditions; however, foundational studies such as this build essential trust and support for future industry collaboration. 

Of the centralized, hierarchical, regional, and decentralized FL architectures, centralized aggregation has demonstrated the best performance \cite{zhang2020federated}, which informs our choice of a central architecture. A range of FL aggregation methods exists, such as FedAvg, clipped FedAvg, momentum, weighted, and hierarchical aggregation \cite{moshawrab2023reviewing}, along with multiple FL implementations, including FedAvg, FedProx, FedBoost, and SCAFFOLD \cite{moshawrab2023reviewing}. This study specifically investigates two methods: (i) FedAvg, as applied to LoRA adapters in a recent SE study on code summarization \cite{kumar2024code}, and (ii) FLoRA, a recent technique that stacks clients' LoRA adapters instead of averaging, as introduced by \cite{wang2024flora}.

FL has been adopted in SE studies recently. For instance, a commit classification study leveraged FL with a BERT-based model for identifying bug-fix commits \cite{shanbhag2022exploring}, while another recent work applied SVM and CodeBERT models for code clone detection and defect prediction \cite{yang2024federated}. A code summarization study used LLaMA2 \cite{touvron2023llama} with LoRA-based FL fine-tuning, demonstrating performance comparable to central models \cite{kumar2024code}. Drawing from this methodology, our study adapts CodeLLaMA \cite{roziere2023code}, a model fine-tuned specifically for code to the task of code translation and investigates both FedAvg and the FLoRA stacking technique for model aggregation. To our knowledge, this is the first work to integrate FL with LLMs for code translation, enhancing data privacy and enabling C\# to Java (and vice versa) translations with a federated approach.

\section{Conclusion and Future Directions}
\label{sec: ConclFuture}
The paper demonstrates a unique approach to automated code translation by applying Federated Learning (FL) to collaboratively fine-tune a Large Language Model (LLM), specifically 7B CodeLLaMA, for Java-to-C\# and C\#-to-Java translation. Our results show that the federated model, created by combining LoRA adapters with both FedAvg and FLoRA aggregation methods, achieves better translation quality than individual client models, and also achieves competitive performance with a centrally trained model while preserving client data privacy.
This study emphasizes FL's potential as an effective, privacy-preserving approach for code translation, encouraging future work to explore FL’s scalability and adaptability to additional programming languages and translation tasks in SE. Future research could further investigate defense mechanisms against compromised or erroneous client models and examine other privacy-preserving strategies applicable to code translation and similar SE tasks.


\balance
\bibliographystyle{ACM-Reference-Format}
\bibliography{references}
\end{document}